\documentclass[twocolumn]{aastex62}

\graphicspath{{./}{Figures/}}
\usepackage{longtable}
\usepackage{amsmath}
\usepackage{mathtools}

\usepackage{hyperref}
\hypersetup{
    pdfnewwindow=true,
    colorlinks=true,
    citecolor=cyan,
    linkcolor=blue,
    urlcolor=blue
}

\newcommand\param[1]{\texttt{#1}}

\def\msun{$M_{\odot}$}

\graphicspath{{./}{Figures/}}

\begin{document}
\title{SDSS DR17: The Cosmic Slime Value Added Catalog}

\correspondingauthor{Matthew C. Wilde}
\email{mwilde@uw.edu}

\author[0000-0003-1980-364X]{Matthew C. Wilde}
\affil{University of Washington, Department of Astronomy, Seattle, WA 98195, USA}

\author[0000-0003-0549-3302]{Oskar Elek}
\affil{University of California in Santa Cruz, 1156 High St, Santa Cruz, CA 95064, USA}

\author[0000-0002-1979-2197]{Joseph N. Burchett}
\affil{University of California in Santa Cruz, 1156 High St, Santa Cruz, CA 95064, USA}
\affil{Department of Astronomy, New Mexico State University, PO Box 30001, MSC 4500, Las Cruces, NM 88001}

\author[0000-0002-6766-5942]{Daisuke Nagai}
\affil{Department of Physics, Yale University, New Haven, CT 06520, USA}

\author[0000-0002-1979-2197]{J. Xavier Prochaska}
\affil{University of California in Santa Cruz, 1156 High St, Santa Cruz, CA 95064, USA}
\affil{Kavli Institute for the Physics and Mathematics of the Universe, 5-1-5 Kashiwanoha, Kashiwa 277-8583, Japan}

\author[0000-0002-0355-0134]{Jessica Werk}
\affil{University of Washington, Department of Astronomy, Seattle, WA 98195, USA}

\author[0000-0002-7327-565X]{Sarah Tuttle}
\affil{University of Washington, Department of Astronomy, Seattle, WA 98195, USA}

\author[0000-0002-8700-7795]{Angus G. Forbes}
\affil{University of California in Santa Cruz, 1156 High St, Santa Cruz, CA 95064, USA}
\affil{Purdue University, 610 Purdue Mall, West Lafayette, IN 47907, USA}

\begin{abstract}The ``cosmic web”, the filamentary large-scale structure in a cold dark matter Universe, is readily apparent via galaxy tracers in spectroscopic surveys. However, the underlying dark matter structure is as of yet unobservable and mapping the diffuse gas permeating it lies beyond practical observational capabilities. A recently developed technique, inspired by the growth and movement of \textit{Physarum polycephalum} `slime mold', has been used to map the cosmic web of a low redshift sub-sample of the SDSS spectroscopic galaxy catalog. This model, the \textit{Monte Carlo Physarum Machine} (MCPM) was shown to promisingly reconstruct the cosmic web. Here, we improve the formalism used in calibrating the MCPM to better recreate the Bolshoi-Planck cosmological simulation's density distributions and apply them to a significantly larger cosmological volume than previous works using the Sloan Digital Sky Survey (SDSS, $z < 0.1$) and the Extended Baryon Oscillation Spectroscopic Survey (eBOSS) Luminous Red Galaxy (LRG, $z \lesssim 0.5$) spectroscopic catalogs. We present the `Cosmic Slime Value Added Catalog' which provides estimates for the cosmic overdensity for the sample of galaxies probed spectroscopically by the above SDSS surveys. In addition, we provide the fully reconstructed 3D density cubes of these volumes. These data products were released as part of Sloan Digital Sky Survey Data Release 17 and are publicly available. We present the input catalogs and the methodology for constructing these data products. We also highlight exciting potential applications to galaxy evolution, cosmology, the intergalactic and circumgalactic medium, and transient phenomenon localization.
\\~
\end{abstract}

\section{Introduction} \label{introduction}

The cosmic web is an emergent prediction of $\Lambda$CDM cosmology and is ubiquitously reproduced and readily identifiable in cosmological simulations, where the underlying density distribution is known \citep[e.g.,][]{Springel:2005_millenium, Vogelsberger:2014zl}.  However, unveiling the large-scale structure in the observational realm using galaxies and absorption lines as tracers of the intergalactic medium (IGM) is much less straightforward. The underlying dark matter distribution remains unobservable. The two most accessible tracers, such as galaxies and quasar absorption lines, are limited by the practical observational constraints of galaxy redshift surveys and the scarcity of quasars in the universe, respectively.  Even when observational tracers are available at relatively high sampling densities, the problem of reconstructing the cosmic web is highly complex. 

We highlight two of the myriad scientific motivations for cosmic web reconstruction. First, of paramount concern in galaxy astrophysics is the impact of a galaxy's environment on its evolution.  Correlations between environmental metrics and galaxy properties, such as morphology \citep[e.g.,][]{Dressler:1980qy}, color \citep[e.g.,][]{Abell:1965fk}, and star formation \citep[e.g.,][]{Balogh:1999xy,peng10}, have been known about for many decades, but the physical mechanisms and their relative importance remain heavily pursued problems.  Galaxy-environment analyses typically fall along one of two paths: local environment-centric or large-scale environment-centric.  In the former, one employs an environmental density metric, such as a nearest-neighbor distance or density within some aperture \citep[]{Kauffmann:2004ai, peng10}, or galaxies are associated with a local group or cluster environment \citep[]{Yang:2007kl, berlind06} and galaxy properties are studied with respect to the properties of the group or cluster \citep[]{Carollo:2013xy, Catinella:2013ys, Poggianti:2009_evolutionInClusters}.

The latter path is less straightforward, as one must infer the large-scale structure from tracers, typically the galaxies themselves, and correlate galaxies back to that structure in some way.  Various methods have been devised to reconstruct the cosmic web from discrete tracers.  \cite{libeskind18} reviewed a number of these, and we refer the reader to this valuable resource for an overview of the techniques employed and comparisons between them.  Once the underlying density field is inferred, one can correlate galaxy properties with this density field (an approach one can directly employ with the catalog described here) or attempt to geometrically relate a galaxy's position to the structure identified, e.g., the distance to a filament.  One should appreciate that filament identification \citep[e.g., DisPerSE;][]{luber19,Tempel:2014aa}, whether from a density field or some other methodology, is a separate problem from the inference of the field itself.  

Studies of galaxy properties and their dependence on the cosmic environment report mixed results.  \cite{kuutma17} find a higher elliptical-to-spiral ratio and decreasing star formation rate (SFR) towards filament spines.  Similarly, \cite{odekon18} report that, at fixed stellar mass, galaxies closer to filaments or in higher density environments are more deficient in HI. These large-scale environmental correlations with galaxies have also been investigated using modern hydrodynamical cosmological simulations. \cite{codis18} measure the spin-filament alignment in IllustrisTNG \citep[]{vogelsberger14} and find a strong dependence on spin alignment with galaxy mass. \cite{pasha22} find that the collapse of large-scale structure into sheets at higher redshifts ($z\sim3$) can create shocks that explain quenching in dwarf galaxies similar to the effects seen in the presence of clusters and groups. 

Second, in addition to the galaxies themselves, the IGM studied in context with the cosmic web environment can yield important insight. \cite{wakker15} measured the Ly$\alpha$ absorption in quasar spectra probing a foreground visually identified filament, finding increasing absorber equivalent width and linewidth with decreasing projected distance to the center of the filament.  With a larger archival sample of QSOs and filaments, \cite{bouma21} find similar results, with Ly$\alpha$ absorbers showing both greater incidence and column density at a small projected distance and velocity offsets from filaments first identified by \citet{Courtois:2013_cosmography}. In the first application of the reconstruction framework we use here, \citet{burchett20} analyzed the Ly$\alpha$ optical depth as a function of cosmic web density probed by QSO sightlines. They found three distinct regimes: (1) a void regime at low matter overdensity with no detected absorption, (2) an onset of absorption in the outer skins of filaments with monotonically increasing optical depth, and (3) the highest-density regime where the absorption no longer increases with local density but rather turns over and declines at the highest densities.  Associating the IGM to the cosmic web provides important constraints on hydrodynamical processes modeled in cosmological simulations that may be used to interpret the environmental quenching conundrums.

In this manuscript, we employ the novel method first introduced in \cite{burchett20} described in detail by \cite{elek2022mcpm}, which is based on the morphology of the \textit{Physarum polycephalum} slime mold organism to map the cosmic density field. This model implicitly traces the cosmic web structure by efficiently finding optimal pathways between the galaxies that trace filaments. We apply our model to two large galaxy catalogs, the NASA Sloan Atlas (NSA) \citep[]{blanton11} and the catalogs of Luminous Red Galaxies (LRGs) from the SDSS-IV Extended Baryon Oscillation Spectroscopic Survey \citep[]{bautista18}. Our method faithfully reconstructs the cosmic matter density of the cosmic web throughout the observed volume, allowing the study of the dark matter distribution with respect to any objects of interest in the survey footprints, not just at the input galaxy locations. We have released this data as part of the SDSS Data Release 17 (DR17) as a Value Added Catalog (VAC) publicly available for the community's use. 

Unless stated otherwise, we adopt the Planck15 \citep{planck16} cosmology as encoded in the \texttt{ASTROPY} package \citep{astropy13, astropy18}.

\section{Data}

We first describe the required inputs for reconstructing the map of cosmic densities produced by MCPM. MCPM takes as input a 3D catalog of galaxy positions with known masses and reproduces a data cube reconstructing the filamentary structure connecting the galaxy halos. To optimize the parameters in MCPM, we also require a known density field from a cosmological simulation to compare our reconstruction. We then apply the tuned model to observational catalogs of galaxies with known masses to reconstruct the physical cosmic web. 

We employ the dark matter-only Bolshoi-Plank $\Lambda$CDM (BP) simulation (described below) as our training density field. We then apply our model to spectroscopic surveys that provide large samples of precise redshifts combined with value-added catalogs that estimate the galaxy masses. We use two primary catalogs for our galaxy positions, the NASA-Sloan Atlas (NSA, or NSA/SDSS) for galaxies with $z < 0.1$ and the Large Scale Structure catalogs from Sloan Digital Sky Survey (SDSS) for galaxies at higher redshifts ($z \lesssim 0.5$). These two catalogs each offer advantages and disadvantages and are described below. Note that no new DR17 data were used in this VAC. We now describe the galaxy catalogs and the simulations used as inputs to MCPM.

\subsection{NASA Sloan Atlas}
The NASA Sloan atlas (NSA) is a value-added catalog constructed from reprocessed SDSS $ugriz$ photometry combined with Galaxy Evolution Explorer (GALEX) photometry in the ultraviolet. It was designed to improve the standard SDSS sky subtraction pipeline \citep[]{blanton11}. We use the most recent version of this catalog, \textbf{nsa\_v1\_0\_1.fits}, which contains galaxies out to $z=0.15$. In order to prioritize completeness in this data set, we we imposed an upper redshift cut to those galaxies with $z=0.1$ resulting in a catalog of 325321 galaxies. We will often refer to this catalog in this paper as simply ``NSA/SDSS" to distinguish it from the other catalogs from BOSS. 

\subsection{LRG catalogs}
For the higher redshift portion of our catalog, we use a sample of Luminous Red Galaxies (LRGs) from the Baryon Oscillation Spectroscopic Survey (BOSS). BOSS was part of the SDSS III project, which at the time of its release, provided the largest survey of galaxy redshifts available in terms of the number of redshifts measured by a single survey and the effective cosmological volume covered. We chose to use the LRG catalogs as tracers of the dark matter (DM) density as these catalogs are more complete at these redshifts with respect the selection function than using the more general SDSS galaxy sample.  The BOSS LRG sample derives from the large scale structure catalogs provided by the team and is broken into Northern and Southern Galactic Cap regions (LRG-NGC and LRG-SGC, respectively) \citep[]{ross11, ho12, ross12}. We use the LOWZ catalogs, which provide a sample of LRGs to $z \lesssim 0.5$ and are found in the files\footnote{ https://www.sdss.org/dr14/spectro/lss/} \texttt{galaxy\_DR12v5\_LOWZ\_North.fits.gz} and \texttt{galaxy\_DR12v5\_LOWZ\_South.fits.gz}. The procedure to create this catalog is mostly based on \cite{reid16} with modifications to the redshift failure and systematic corrections described in \cite{bautista18}.

\begin{figure*}[t] 
\includegraphics[width=\textwidth]{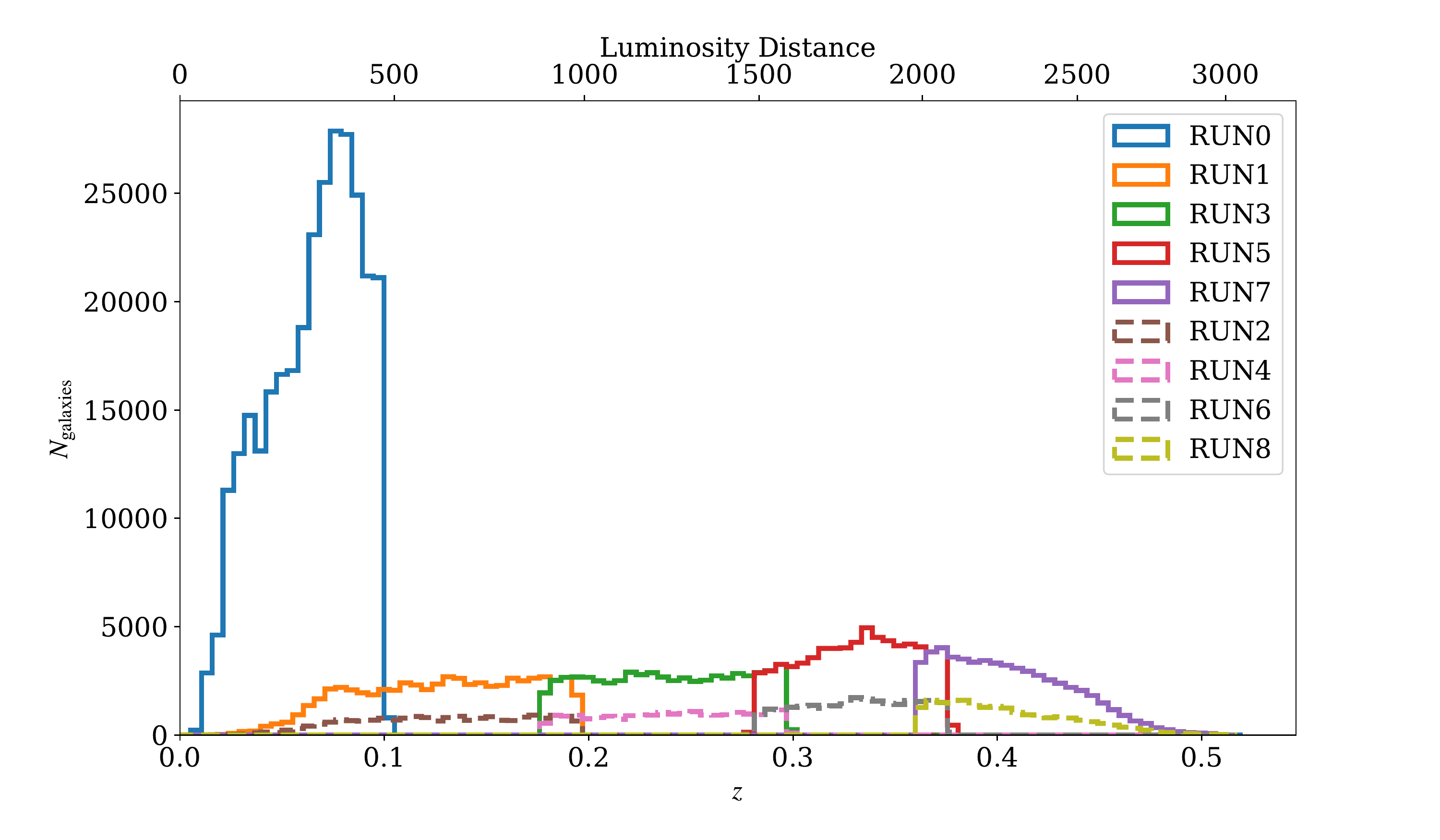}
\caption{Distribution of the galaxy redshifts for the NSA/SDSS (blue, solid) and LRG-NGC (multi-colored, solid) and LRG-SGC (multi-colored, dashed) data sets that were used to reconstruct the cosmic density map. The NSA/SDSS catalog includes all galaxies out to $z=0.1$ and is denoted as RUN0 in the MCMP VAC. The LRG catalogs extend to higher redshifts but only include the rarer LRGs, hence the lower galaxy count. This figure also shows the slicing scheme used to self-consistently fit the MCPM model in subsets of redshift as the density of galaxies decreases with luminosity distance in comoving Mpc.
~\\
\label{fig:histogram}}
\end{figure*}

\begin{figure*}[t] 
\includegraphics[width=\textwidth]{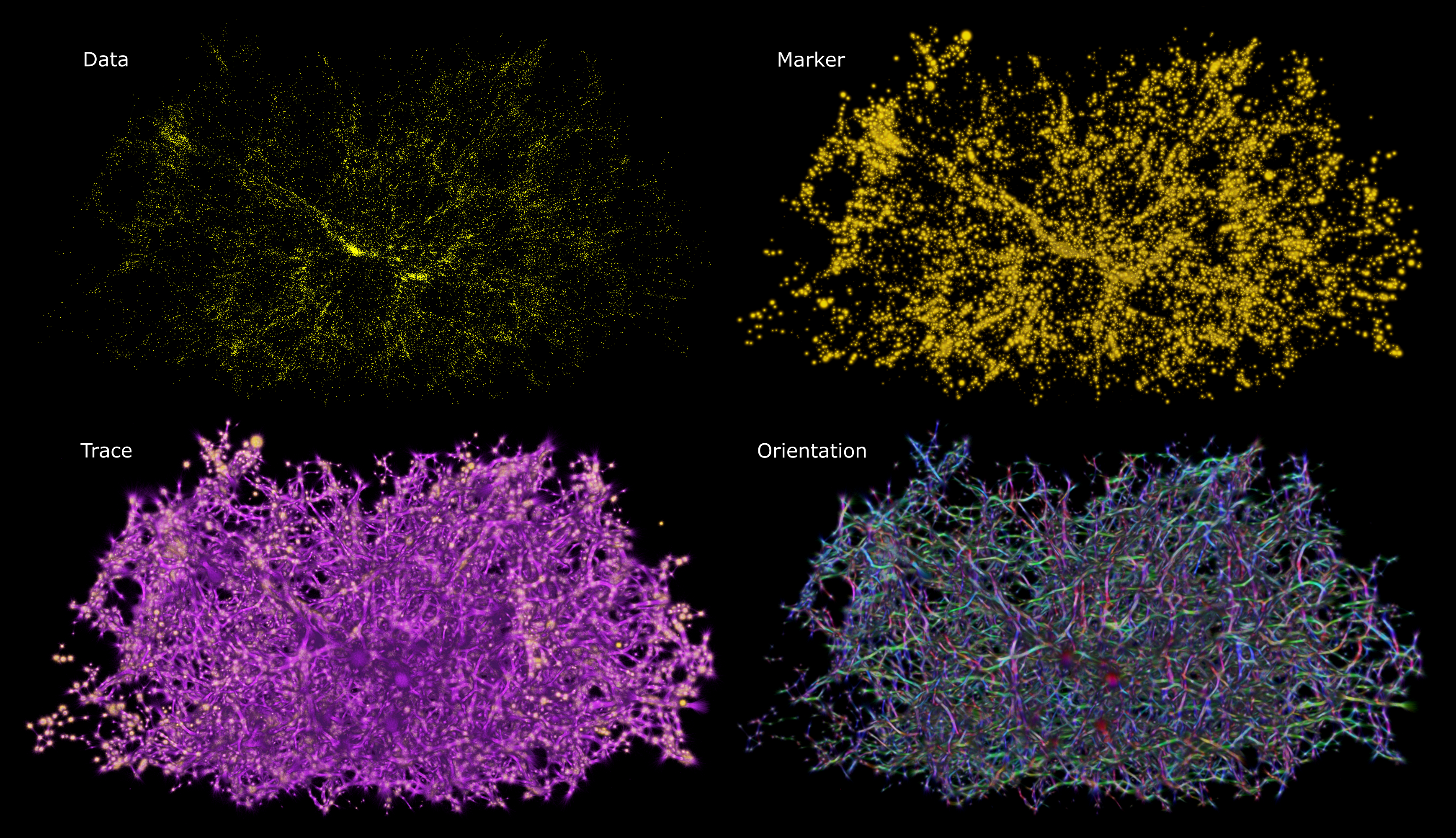}
\caption{Overview of MCPM's operating modalities, demonstrated on the $0.018 < z < 0.038$ sample of SDSS galaxies. Clockwise from top left: input data points and the marker concentration emitted by the data (yellow), reconstructed trace field $f_\mathrm{T}$ (purple), corresponding orientation field $f_\mathrm{O}$ (XYZ directions mapped to RGB colors).
~\\
\label{fig:mcpm-modalities}}
\end{figure*}

\begin{figure*}[t]
\includegraphics[width=\textwidth]{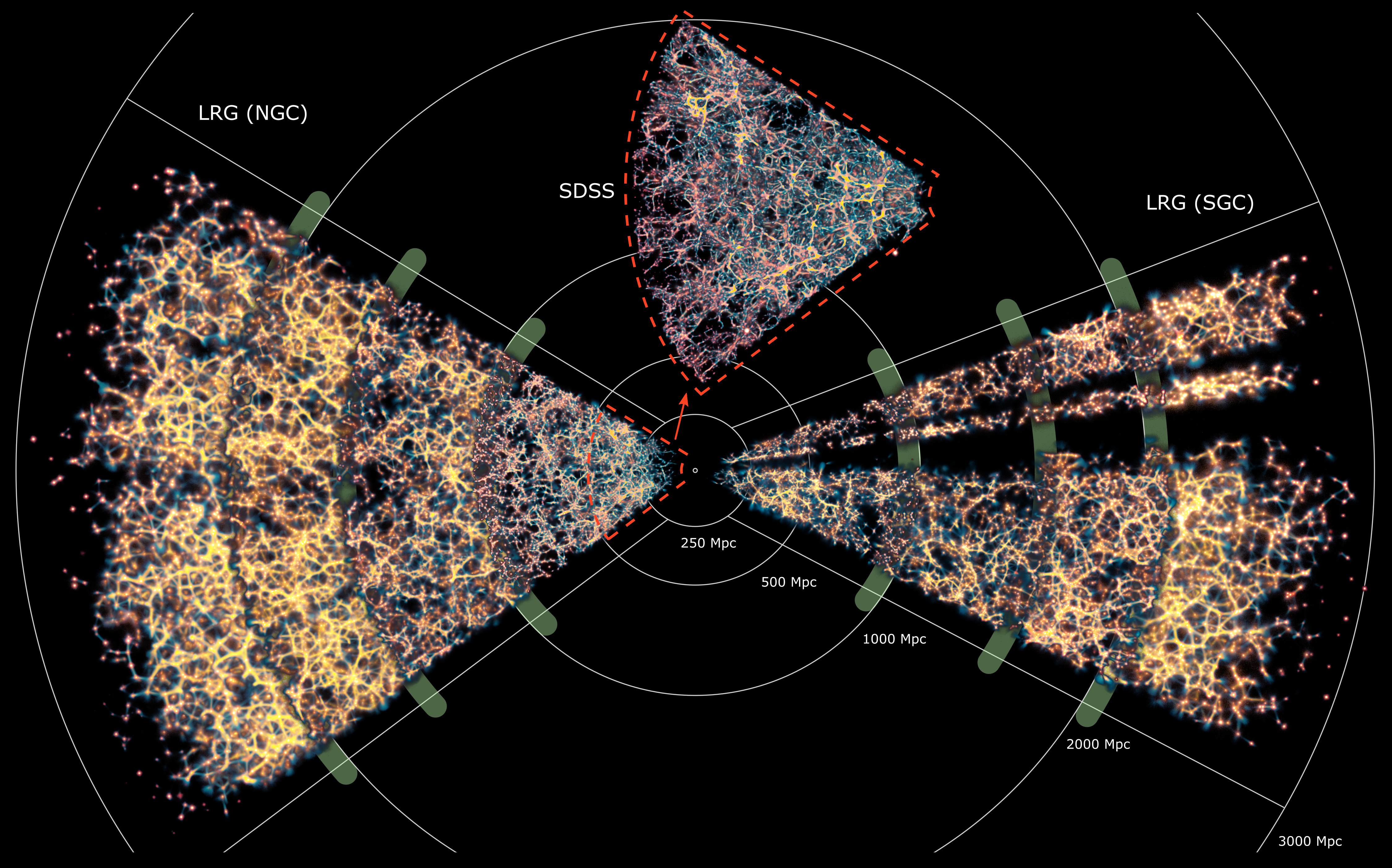}
\caption{Our reconstructed cosmic web data products and their spatial relation to another. The green bands highlight regions of overlapping LRG slices. The SDSS portion of the data is magnified to visualize the higher amount of recovered structure owing to the denser observations.
~\\
\label{fig:slices-diagram}}
\end{figure*}

\subsection{Mass Determination}

We used the LRG galaxy stellar masses from the Firefly VAC. The Firefly VAC\footnote{https://www.sdss.org/dr16/spectro/eboss-firefly-value-added-catalog/} \citep{comparat17} provides galaxy properties of all SDSS, BOSS, and eBOSS spectra using the FIREFLY fitting routine \citep{wilkinson17} (v1\_0\_4 for DR14 and v1\_1\_1 for DR16), which incorporates the stellar population models of \cite{maraston11}. The Firefly catalog includes light- and mass-weighted stellar population properties (age and metallicity), E(B-V) values, and most crucially to this work, stellar mass for all galaxies in the catalog. We used the DR14 catalog to determine masses for the galaxies in the file \texttt{sdss\_eboss\_firefly-dr14.fits}.

The lower redshift NSA/SDSS catalog contains many galaxies that are spatially resolved and require more careful photometric analysis \citep[e.g.,][]{blanton11}. The most recent version of this catalog provides elliptical Petrosian aperture photometry, which is more accurate than the standard SDSS pipeline. We adopt the Petrosian aperture-derived mass to estimate the galaxy's stellar mass for this sample. 

\subsection{Bolshoi-Planck Simulations}
To calibrate our MCPM density estimates to the cosmic matter density, we use the dark matter only Bolshoi-Plank $\Lambda$CDM (BP) simulation \citep[]{klypin16, rodriguez-puebla16}. The BP simulation uses 2048$^3$ particles in a volume of 250h$^{-1}$ Mpc$^3$ and is based on the 2013 Planck \citep{planck14} cosmological parameters and compatible with the Planck 2015 parameters \citep{planck16}. We utilize density field from the simulation smoothed by Gaussian kernel over scales of 0.25 Mpc h$^{-1}$ \citep[]{lee17, goh19}. We also employ the BP halo catalog produced using the \texttt{Rockstar} algorithm \citep[]{behroozi12}.

\section{Methodology}

\subsection{The MCPM algorithm}
We produced the VAC data with the Monte Carlo Physarum Machine (MCPM) algorithm implemented in the \textit{Polyphorm} software\footnote{https://github.com/CreativeCodingLab/Polyphorm}. MCPM was first used in \cite{burchett20} to reconstruct a 3D density field estimate of the large-scale structure spanning 37.6k SDSS galaxies within the $0.018 < z < 0.038$ range. The detailed description of methodology and analyses are described in \cite{elek2022mcpm}. We provide a brief summary of the model here.

MCPM is a massively parallel agent-based model inspired by the growth patterns of \textit{Physarum polycephalum} slime mold. Its main modalities are visualized in Figure~\ref{fig:mcpm-modalities}. Using a swarm of millions of particle-like agents, MCPM iteratively traces the network structures implicit in the input data: dark matter halos or galaxies represented as a weighted 3D point cloud. In linear proportion to their halo mass, the data points emit a virtual marker which the agents navigate toward at every iteration.

The key innovation of this model is the probabilistic navigation of the agents: the sampling of their trajectories according to PDFs derived from the data-emitted marker field. For reference, the deterministic baseline model, where the agents always follow the maximum marker concentration, leads to the collapse of some filamentary configurations and the omission of a significant portion of data points, approximately a third as measured in \cite{burchett20}. In contrast, MCPM fits over 99\% of all input data points and can reconstruct configurations where multiple filaments branch out from a single origin, e.g., in massive galaxy clusters.

MCPM produces two main quantities: the \textit{trace} field and the \textit{orientation} field. The trace field $f_\mathrm{T}: \mathbb{R}^3 \rightarrow \mathbb{R}_+$ accumulates the superimposed trajectories of all active agents and represents the reconstructed LSS density field (after statistical standardization, Section~\ref{sec:standardization}). The orientation field $f_\mathrm{O}: \mathbb{R}^3 \rightarrow \mathbb{R}_+^3$ records the averaged unsigned directions of the agents and serves as a clustering criterion in our FoG compensation step (Section~\ref{sec:FoG-compensation}). Both are robust (i.e., stable in time) Monte-Carlo estimates of the equilibrium agent distributions.

Compared to our earlier applications of the MCPM model \citep[]{burchett20,simha20}, we introduce a few methodological and implementation changes aimed at improving the quality of the fits (more on this in Section~\ref{section:fit_to_sim}):
\begin{enumerate}
    \item Linear accumulation of $f_\mathrm{T}$ and $f_\mathrm{O}$ values instead of the original exponential floating window averaging. The latter is used for the supervised part of the fitting when exploring different MCPM configurations. After finding the optimal data-specific set of model parameters, we switch to linear averaging, which dramatically reduces the solution variance.
    \item To avoid numerical errors, we increase the numerical precision from \param{fp16} to \param{fp32} for both $f_\mathrm{T}$ and $f_\mathrm{O}$. This slows the implementation by 10-20\%, which is acceptable for maintaining interactivity during fitting.
    \item We redesigned the agent rerouting step. Rerouting is invoked when an agent encounters no data for too many subsequent steps, indicating either a boundary of the dataset or a large void. Our original rerouting assigned such an agent to a random location in space; currently, we repositioned it to the location of a random data point. This change leads to a significant decrease of background noise and effectively increases the dynamic range of the obtained solutions for both $f_\mathrm{T}$ and $f_\mathrm{O}$.
\end{enumerate}

\subsection{MCPM fit to Bolshoi-Planck}\label{section:fit_to_sim}

This section describes how we calibrate the MCPM algorithm using the Bolshoi-Planck data. We refer readers to \cite{elek2022mcpm} for more details of the fitting procedure and the impact of the model hyperparameters on the resulting reconstruction geometry. Readers interested in the catalog data can skip to Section~\ref{sec:SDSS-fit}.

Fitting MCPM to input data (either galaxies or halos) is a semi-supervised procedure, where the operator focuses on maximizing the fitness function $E$ while maintaining the connectedness and continuity of the reconstructed geometry. We define the fitness $E$ of a given reconstructed trace field $f_\mathrm{T}$ over a dataset $D$ as
$$
E(f_\mathrm{T},D) = \frac{1}{|D|} \sum_{d \in D} \frac{f_\mathrm{T}(d_{\mathrm{position}})}{d_{\mathrm{mass}}}.
$$
This results in a maximum likelihood estimator normalized by each data point's mass to avoid overfitting to the most massive objects (given the large dispersion of typical galaxy and halo masses). Since we do not yet have a precise mathematical description of the fit's connectedness, we rely on the interactive visualization in \textit{Polyphorm} to ensure that the fit does not collapse into a disconnected set of `islands'. Defining this property rigorously and developing a fully automated fitting procedure remains a future work for us.

To calibrate MCPM's hyperparameters, we fit the model to two snapshots of the Bolshoi-Planck simulation dataset (at $z=0$ and $z=0.5$, both containing roughly 16M halos extracted with the Rockstar algorithm). We adopted some of the parameter estimates from our previous work \citep[]{burchett20}, including the \param{sensing angle} at 20 deg, \param{moving angle} at 10 deg, \param{moving distance} at 0.1 Mpc and \param{persistence} of 0.9 (now adjusted to 0.92 due to the finer granularity of halos used here). We focused on constraining the remaining critical parameters: \param{sampling exponent} (which controls the acuity of obtained structures, especially filaments) and \param{sensing distance} (which determines the scale of the structures, such as mean segment length and by transition the diameter of loops, voids, etc). In addition, we maximize the \emph{monotonicity} of the obtained overdensity mapping as shown in Figure~\ref{fig:bestfit} as an additional constraint when determining the \param{sampling exponent}.

Using this fitting procedure, we matched the MCPM fits to the ground truth densities in Bolshoi-Planck. We determined the optimal \param{sampling exponent} to be 2.5 at $z=0$ and 2.2 at $z=0.5$, which is consistent with the observation that the LSS at higher redshifts is less condensed. For the \param{sensing distance}, the optimal value was 2.37 Mpc. It is worth noting that these \param{sampling exponent} and \param{sensing distance} values pose lower limits for the values used to fit the observational data, because of the significantly lower spatial density of data points in the galaxy catalogs relative to BP simulations, which was compensated for by proportionally increasing the two parameters.

\begin{figure*}[ht!]
\vspace{-1.5cm}
\centering
\includegraphics[width=0.9\textwidth]{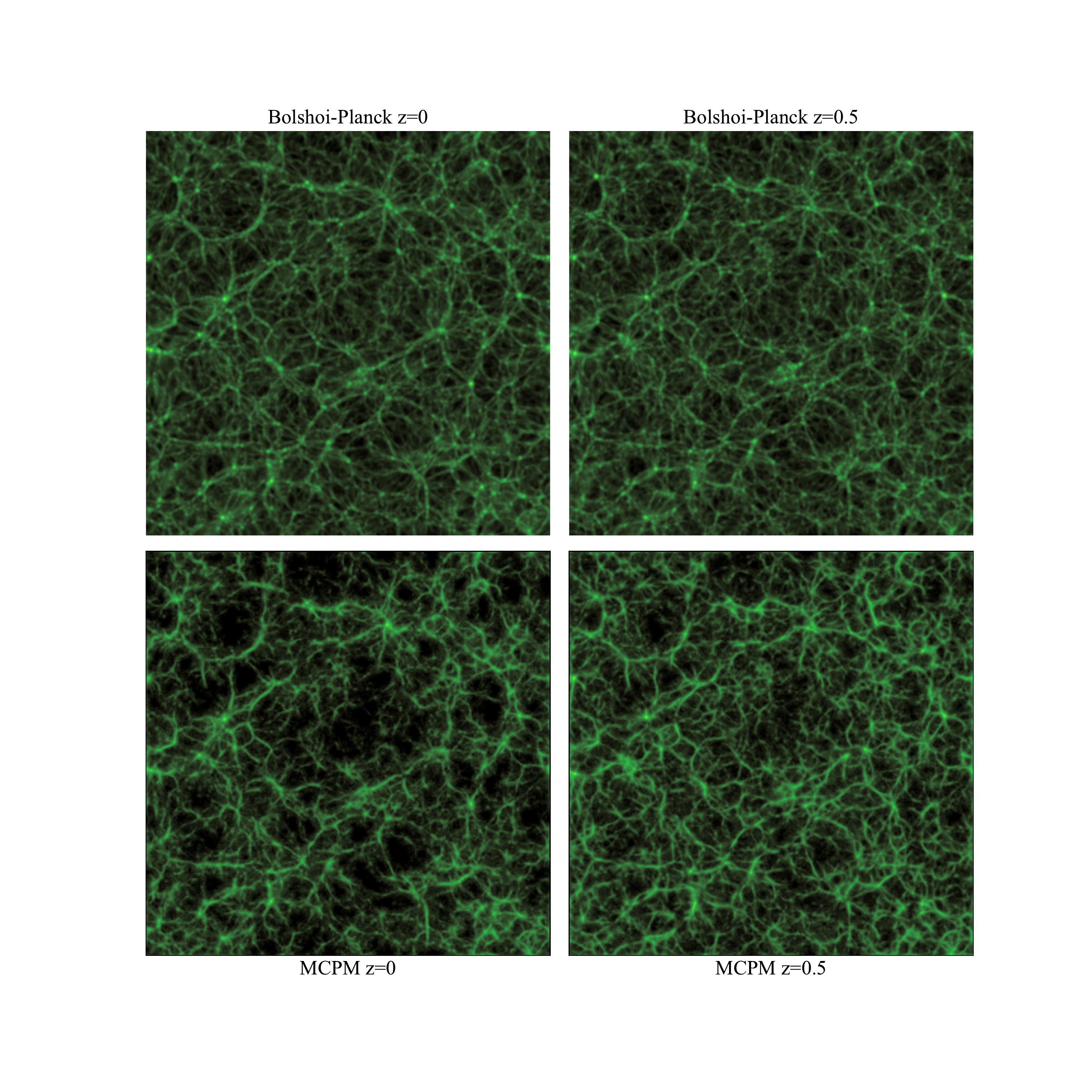}
\vspace{-1.5cm}
\caption{Comparison of the Bolshoi-Planck simulations (top row; where the density field is known) at redshifts of $z=0.0$ (left) and $z = 0.5$ (right) to the MCPM trace of the simulations (bottom row; density recovered from halos alone). MCPM faithfully reconstructs the cosmic structure from the galaxy halo population.
~\\
\label{fig:mcpm_bp_redshift}}
\end{figure*}

\begin{figure}[ht!]
\vspace{-1.2cm}
\centering
\includegraphics[scale=0.36]{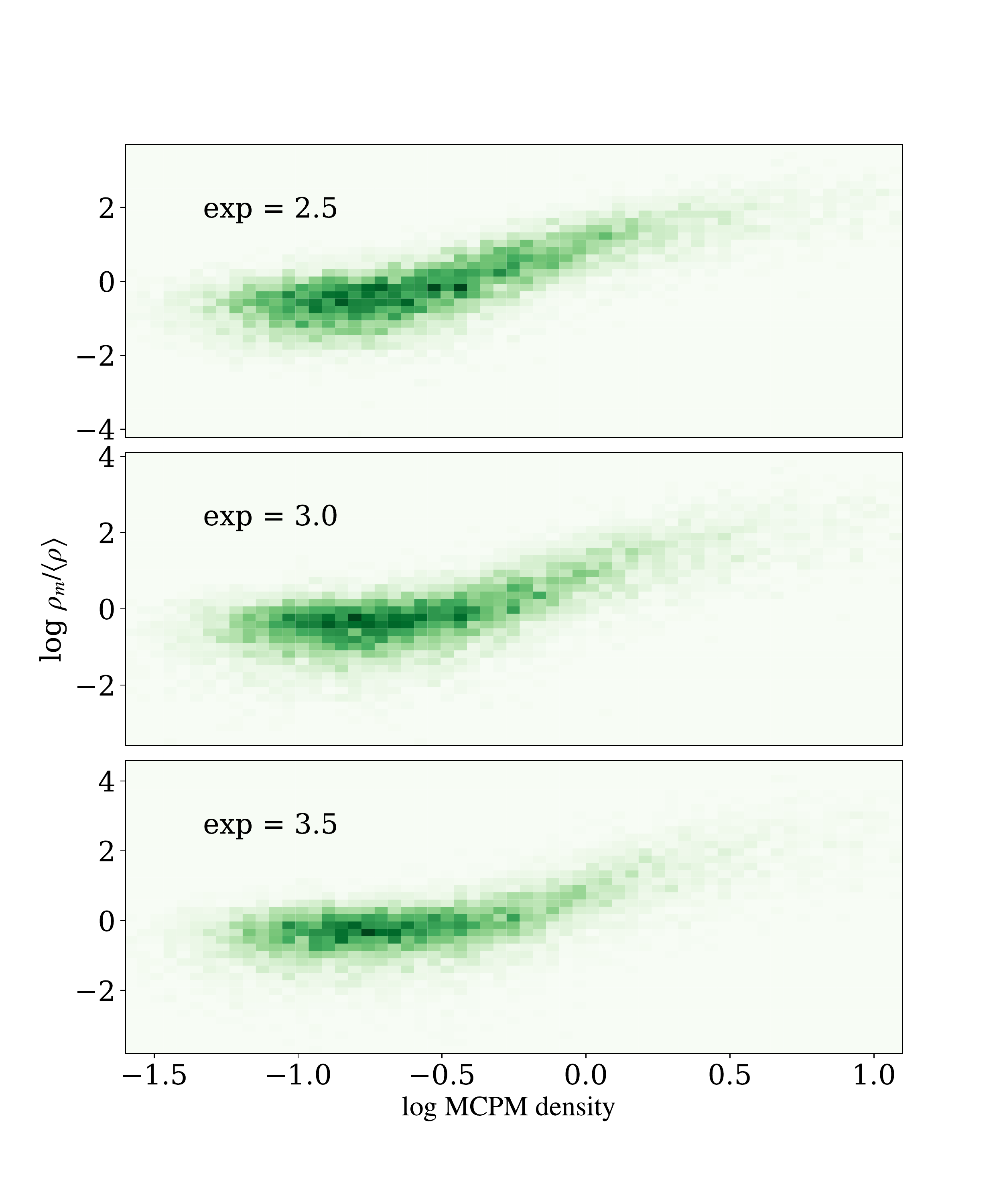}
\vspace{-1cm}
\caption{Comparison of different sampling exponents in increasing order from top to bottom. We find that a sampling exponent of 2.5 produces the most linear mapping between the MCPM densities and the cosmic matter densities from the BP simulations, especially at lower densities where previous versions of the MCPM have generally failed to recover the lowest density structures. \citep[see Figure~10 in][]{burchett20}.
\label{fig:bestfit}}
\end{figure}

In Figure~\ref{fig:mcpm_bp_redshift}, we demonstrate that MCPM reconstructs not just the halos that we feed into it but the cosmic structure, including filaments and voids. More quantitative assessments are available in \cite{elek2022mcpm}.

\subsection{Fit to NASA-Sloan Atlas}\label{sec:SDSS-fit}

The first component of the VAC is based on the MCPM fit to the NASA-Sloan Atlas catalog for $0<z<0.1$, which contains roughly 325k galaxies in luminosity distances between 44 and 476 Mpc. Similar to the BP dark matter halos, we treat the galaxies as 3D point attractors, in this case, weighted by their stellar masses.

The fits are based on the hyperparameters calibrated on the BP simulations. Furthermore, to reflect the lower spatial density of the galaxies in comparison to the halos, we adjust the two critical parameters of MCPM: \param{sampling exponent} to 3.5 and \param{sensing distance} to 5.2. To make these adjustments, we again used the semi-supervised fitting procedure described in Section~\ref{section:fit_to_sim}.

To verify the consistency of the fit across different $z$ values, we have split the SDSS catalog into 3 overlapping slices (44-270 Mpc / 250-370 Mpc / 350-476 Mpc, each containing about 120k galaxies) and fitted them separately by only adjusting the \param{sensing distance} parameter. The resulting optimal values (Figure~\ref{fig:slices}) follow a linear trend, implying that the spatial density of galaxies decreases in corresponding proportion. However, the obtained variation of \param{sensing distance} (3.8--5.6) is well within the ability of the model to perform a consistent fit using a single parameter value. Therefore, we opt for a single fit to the entire catalog using the aforementioned \param{sensing distance} value of 5.2.

\subsection{Fit to LRG Catalogs}

The procedure of fitting to the LRG NGC and SGC catalogs is identical with the SDSS data: using the \param{sampling exponent} of 3.5 and the BP-calibrated values for the other hyperparameters, we continued increasing \param{sensing distance} until reaching an optimal fit.

\begin{figure}[tb] 
\includegraphics[width=\columnwidth]{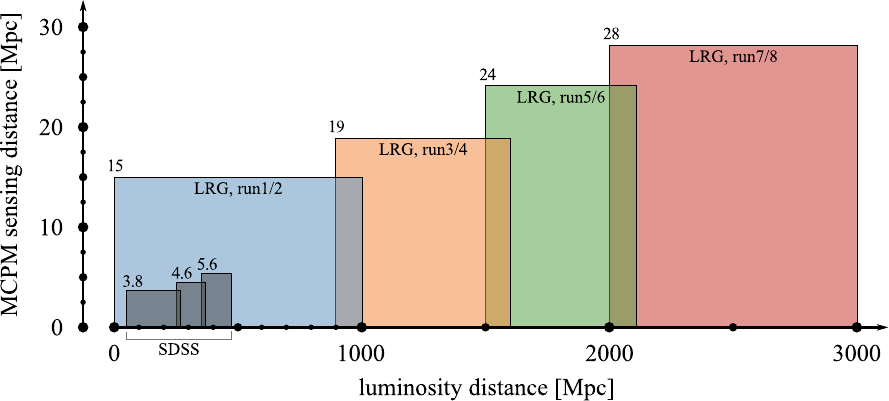}
\caption{Plot of MCPM agents’ \param{sensing distance} (the main feature scaling parameter) as read out from the best fits for the LRG data, radially sliced into 4 runs at overlapping luminous distance intervals. For comparison, we also show the best-fit sensing distances for 3 SDSS slices, which manifest a similar linear growth as we observe in the LRG data.
\label{fig:slices}}
\end{figure}

Due to the much lower spatial density of LRG observations compared to SDSS, the optimal values of \param{sensing distance} end up being considerably higher (Figure~\ref{fig:slices-diagram}). Also, unlike SDSS, the LRG galaxies span a significantly more extended range of redshifts. The consequence is nearly a two-fold growth of the optimal \param{sensing distance} value across the catalog's redshift range (Figure~\ref{fig:slices}). Therefore to construct the VAC, we split the LRG galaxies into 4 overlapping `slices' of approximately equal numbers of galaxies (about 70k per slice for NGC, about 25k per slice for SGC) and fit each separately. The resulting distance intervals are 0-1000 Mpc ($z\approx0-0.2$), 900-1600 Mpc ($z\approx0.18-0.3$), 1500-2100 Mpc ($z\approx0.28-0.38$), and 2000-3000 Mpc ($z\approx0.36-0.51$).

Figure~\ref{fig:slices-diagram} shows the visualization of all obtained density slices and their spatial relations. An added benefit of this approach is the higher resolution of each slice we can afford. This is desirable again due to the massive redshift range of the LRG data.

\subsection{Statistical Standardization \& Mapping}\label{sec:standardization}
The MCPM densities that fit each survey slice, although related to the true physical density, are rather the density of agents in the fit. To translate the MCPM density to cosmic overdensity, we standardize each distribution to the MCPM fit of the simulation so that a mapping between MCPM and cosmic overdensity can be applied. The MCPM fits to the galaxy surveys differ from the fits to the BP simulations because they suffer from luminosity selection functions and are thus much sparser. This particularly affects the lowest density regime of the density distribution. To account for this effect, we used the Wasserstein distance\footnote{https://docs.scipy.org/doc/scipy/reference/generated/\\scipy.stats.wasserstein\_distance.html} or the ``Earth Movers Distance" to calculate the \textit{stretch} and \textit{shift} values such that the distribution of MCPM densities of the surveys could be linearly transformed to best fit the BP-MCPM fit. That is \texttt{TargetDist} $=$ \textit{stretch}$\times$\texttt{SurveyDist} $+$ \textit{shift}. Where the \texttt{TargetDist} is the BP-MCPM density distribution, and \texttt{SurveyDist} is the density distribution of each survey slice. The benefit of this method is that we can impose a lower limit on the density distributions to only take into account the higher density wing of the distribution corresponding to densities that contain structure and avoid the empty space in the survey fits. 

\begin{figure}
%

\centering
\vspace{-0.5cm}
\includegraphics[width=0.51\textwidth, trim=1cm 3cm 0 0]{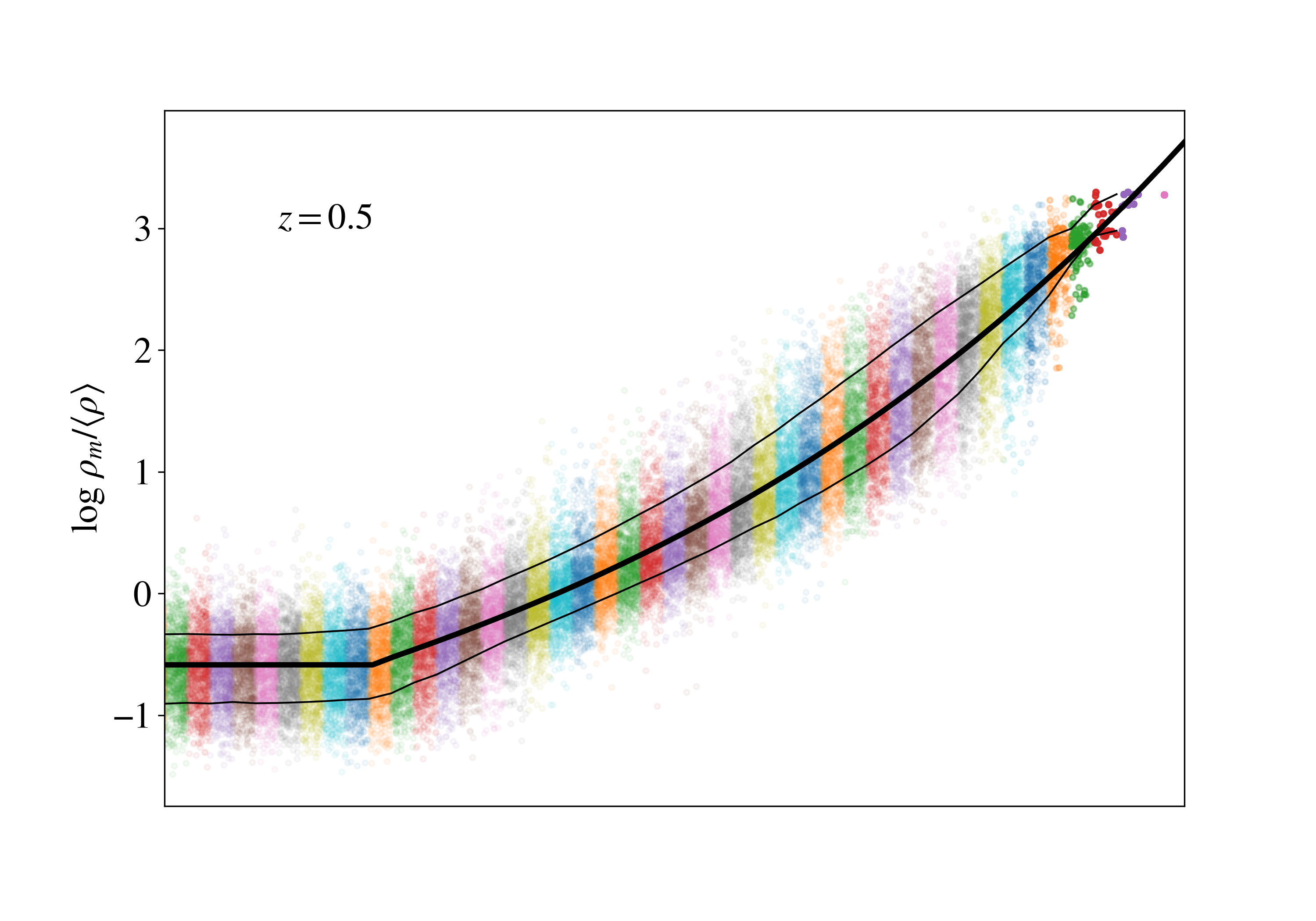}\\%
\includegraphics[width=0.51\textwidth, trim=1cm 0 0 0]{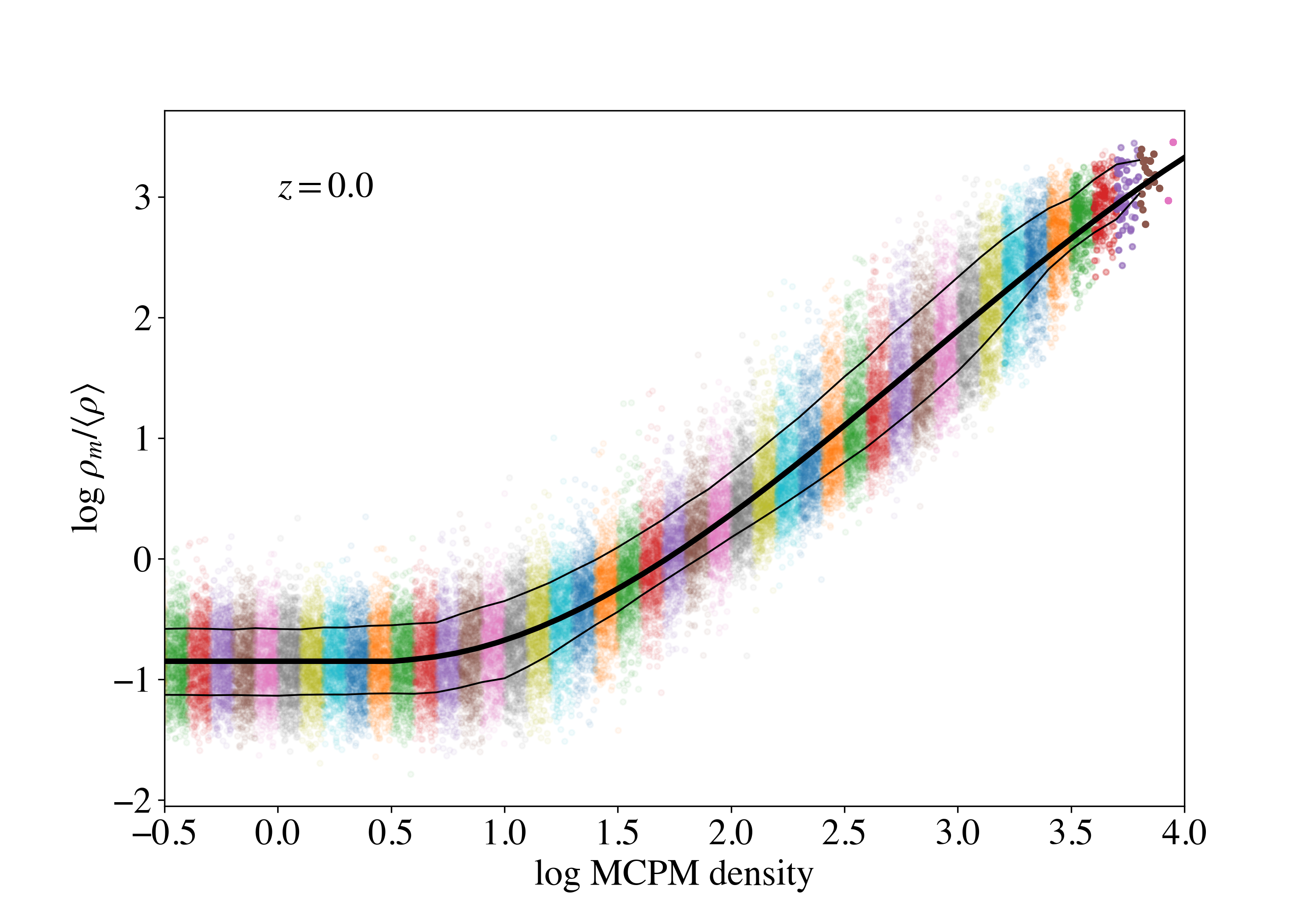}
\vspace{-0.4cm}
\caption{Mapping of the MCPM derived density to the cosmic matter density from the BP simulation. The MCPM densities were binned evenly in MCPM space in bins of 0.1 dex as demarcated by the colored stripes. The custom ReLU mapping function fit to the medians of the bins (thick black line) and $1\sigma$ limits (thinner black lines) are plotted on top of the data. This mapping function provides a translation from the MCPM density to the cosmic overdensity.
}
\label{fig:mapping}
\end{figure}

In order to retrieve the cosmic matter density, $\rho_m / \langle \rho_m \rangle$, we must map the MCPM trace density to that of the BP simulations at each redshift. We fit the BP simulations using the MCPM algorithm and then apply a mapping from MCPM density to cosmic matter density. This mapping was achieved by sampling the MCPM fits in bins of equal density and then determining the density from the BP simulations at the same location. This is shown by the multi-colored stripes in Figure~\ref{fig:mapping}. We then determine the median (and $1\sigma$ limits) of each MCPM density bin. The median densities in each bin were then used to create a mapping function. We based our mapping function on the rectified linear activation function (ReLU), where the maximum change of the median of the bins determines the inflection point. On the right-hand side of the flat part of the function, we fit a cubic polynomial to the data, creating a piece-wise continuous mapping function. This method was chosen over other methods, such as interpolating the bins or using a spline function because fitting to densities above or below those found in the MCPM fits is not well defined. Our method is illustrated in Figure~\ref{fig:mapping} where the thicker black line shows the mapping function applied to the $z=0$ simulation. The thinner black lines show the $1\sigma$ limits of our mapping, which correspond to $\pm 0.5$ dex in log cosmic matter overdensity, $\rho_m / \langle \rho_m \rangle$. 

\subsection{Correction for Redshift Space Distortions}\label{sec:FoG-compensation}

As MCPM operates in 3D space, applying the algorithm necessitates attaching physical distances to the input dataset.  Although distance measurements via more direct methods (e.g., tip of the red giant branch or Type Ia supernovae) \citep[]{Tully:2016wo} may be available for a small subset of the galaxies (and therefore tracers of the underlying density field), we must primarily assume distances concordant with the Hubble flow.  Thus, we initially attach to each galaxy the luminosity distance given the adopted cosmology and galaxy redshift.  Denser environments such as galaxy groups and clusters will include galaxies with large peculiar velocities.  These peculiar velocities will result in redshift space distortions (RSDs), or `fingers of god' (FoG), if adopted directly.  For example, a typical velocity dispersion for a $>10^{14}$ \msun~galaxy cluster ($\sim$ 1000 km/s) would propagate to a systematic error in the distance by assuming pure Hubble flow of $>10$ Mpc.  This issue plagues our low-redshift SDSS sample significantly more than the LRG samples for two reasons: 1) Low-mass galaxies are much more abundant and likely to be observed at low $z$ in the magnitude-limited SDSS, which results in many objects composing apparent false structures along the direction pointing away from (and towards) the observer; (2) High-mass galaxies, which will dominate the samples at progressively higher redshifts, preferentially reside as central galaxies in their local environments \citep{Lan:2016lr}.  Therefore, these galaxy samples will be less subject to systematic error in cosmological distance than our lowest redshift sample.  Thus, we employ an RSD correction for the $z<0.1$ SDSS galaxy sample that we detail here.  

\begin{figure*}[tb] 
\begin{minipage}{\linewidth}
\centering
\includegraphics[width=0.9\linewidth]{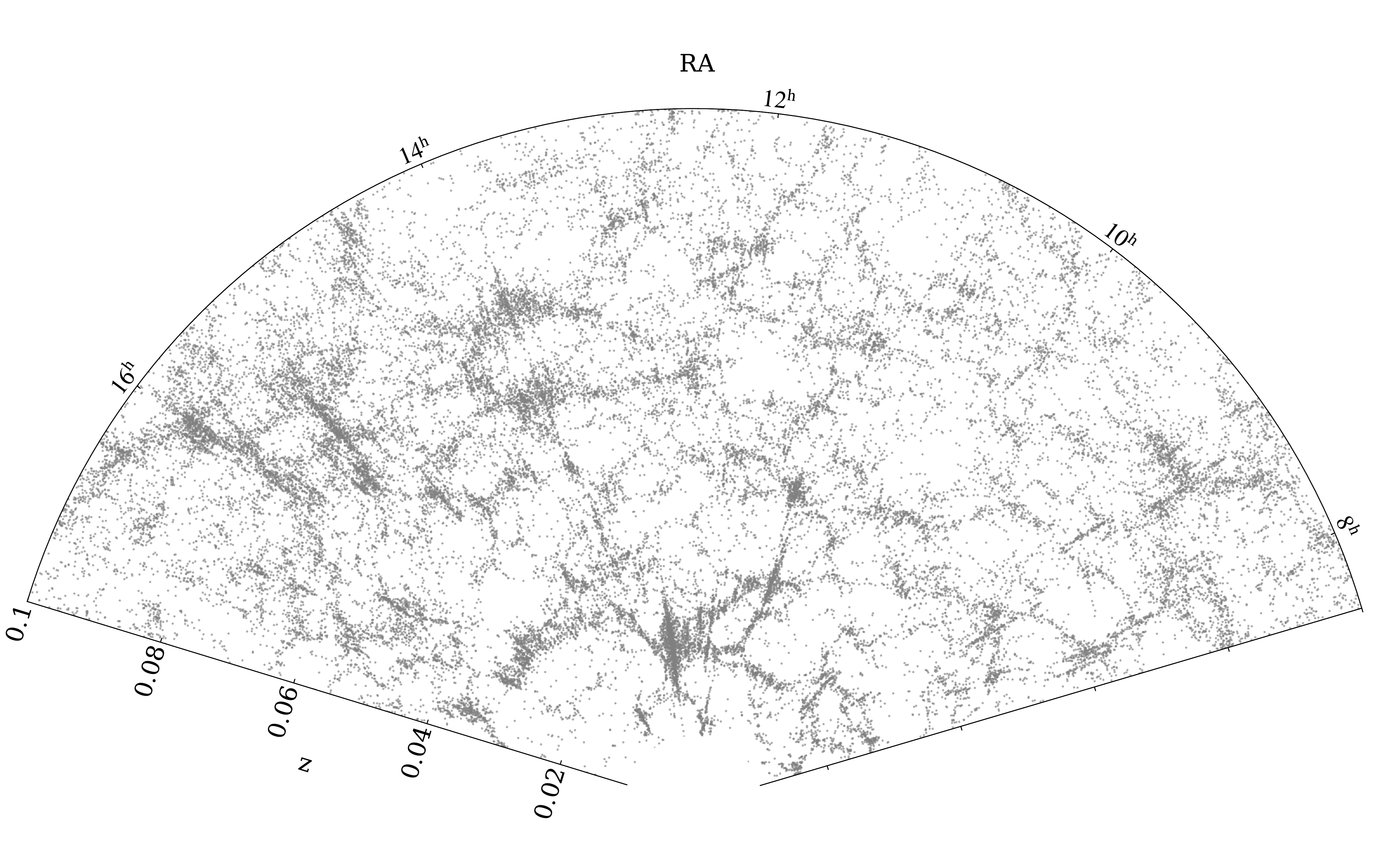}

\end{minipage}
\begin{minipage}{\linewidth}
\centering
\includegraphics[width=0.9\linewidth]{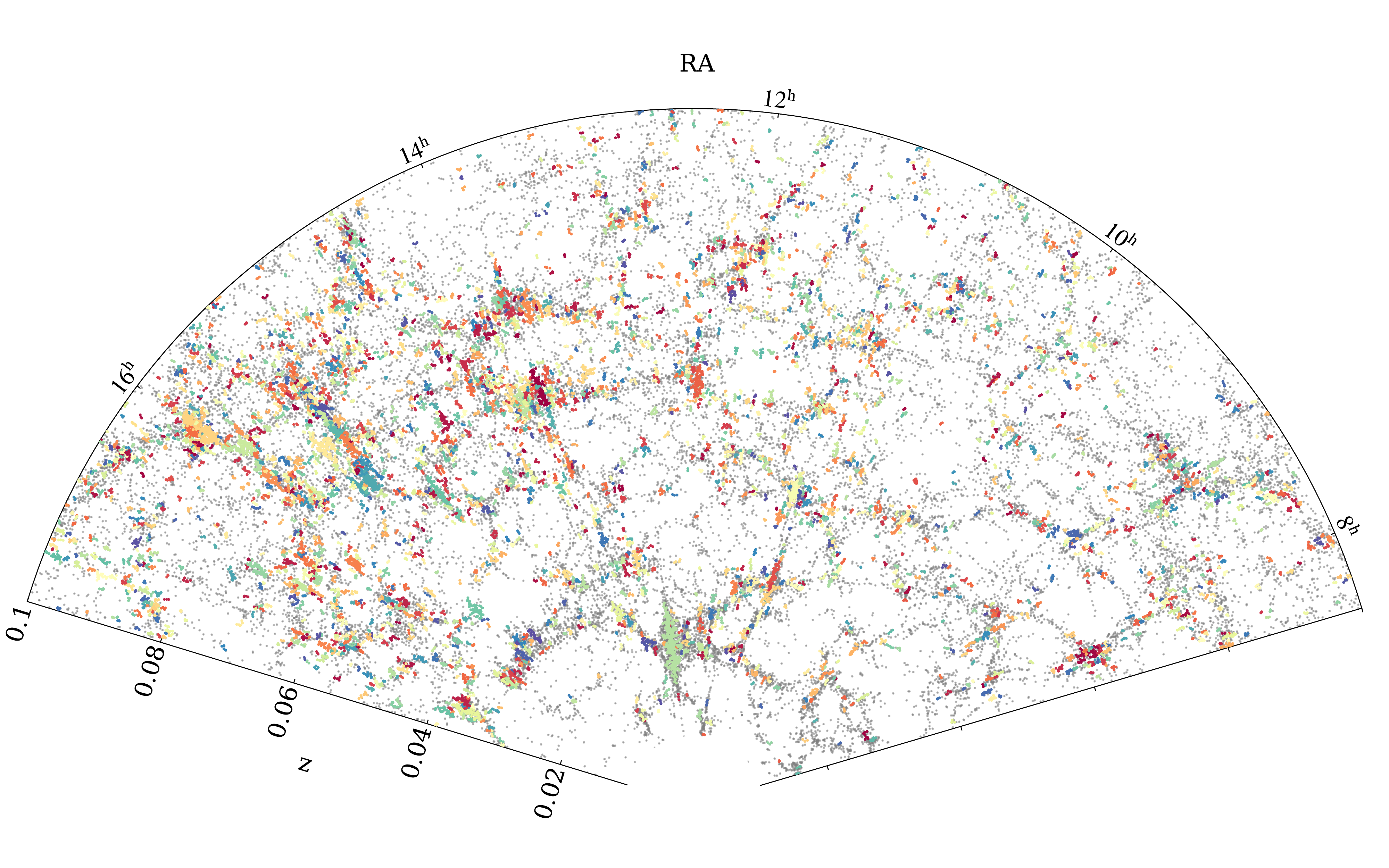}
\caption{A slice in declination of our input galaxy catalog (grey points, top). RSD structures identified by DBSCAN are shown in various colors overlayed on the original points (bottom).\label{fig:FoGs_color}}
\end{minipage}
\end{figure*}

A key feature of MCPM is that the cosmic web reconstruction converges to an equilibrium state but is a dynamical system nonetheless. The adopted `densities' are aggregated trajectories of the millions of agents seeking efficient pathways between galaxy tracers.  MCPM also outputs the components of an aggregated three-dimensional agent velocity vector for each cell in the volume.  We use these velocities to identify RSDs, as the agent velocities producing them will be preferentially oriented perpendicular to the plane of the sky along the line of sight and will be clustered in their celestial coordinates.  We select points in the MCPM cube by orientation as follows: we (1) convert each input galaxy's location in the MCPM-output cube to its equivalent celestial coordinates, (2) find the three components of a unit radial vector parallel to the line of sight in Cartesian space to match the coordinate system of the MCPM velocity vectors, and (3) calculate the dot product between the aggregated velocity vector at each galaxy's position in the cube with the unit radial vector and assign the result to that galaxy.  Galaxies within an RSD structure (FoG), having either parallel or antiparallel velocity vectors to the unit radial vector, should not have dot product absolute values close to zero.  Therefore, we filter out galaxies with dot product absolute values less than 10, chosen upon inspecting the distribution of galaxy dot product values as a conservative cut.   To identify galaxy positions with similar velocity orientation \textit{and} projected location on the sky, we then employ the Density-Based Spatial Clustering of Applications with Noise (DBSCAN) algorithm as implemented in the \texttt{scikit-learn}\footnote{https://scikit-learn.org/stable/} python package, feeding it the sky coordinates and redshift.  For this step, we further filter the galaxy catalog by mass to those with $M_* > 10^{10}$ \msun, as the completeness of SDSS declines for less massive galaxies at the upper end of our redshift range $(z\sim 0.1)$.  DBSCAN operates by locating high-density cores in the data, which are the beginnings of the clusters. The algorithm searches out from these cores, adding points until no more points are found in within some distance tolerance (in whatever space the data occupy).  This algorithm contains several advantages over other possible choices, including scalability, compatibility with non-flat geometries, and the feature that certain points may not be included in any cluster (they are deemed `noise'). Two critical parameters for DBSCAN are the distance tolerance (\texttt{eps}) and the minimum number of points to be considered a core in the data (\texttt{min\_samples}). We chose \texttt{min\_samples}=3 as a minimum number of galaxies (e.g., such as in a group or cluster) that might form a false RSD structure (FoG) in the MCPM model.  We chose a value of \texttt{eps}=2 upon experimenting with several values through visual inspection to balance the inclusion of FoGs (which are readily identified by the eye), containing a relatively small number of galaxies while minimizing false identification of filaments not oriented antiparallel to the plane of the sky as RSD structures. Figure~\ref{fig:FoGs_color} shows the resulting clusters identified by DBSCAN in a slice in declination of our galaxy catalog, with galaxies belonging to the same cluster having the same color.  

From the output clusters identified by DBSCAN, we find the velocity range spanned by galaxy redshifts within each cluster using the full-width-half-maximum (FWHM) of the velocity distribution ($v_{\rm FWHM}$).  For clusters with $v_{\rm FWHM} >$ 300 km/s, we adopt new redshifts for the associated galaxies to be commensurate with more realistic physical distance separations inferred from a simple luminosity distance based on the redshift; this procedure is as follows.   Assuming the cluster members are bound to the same virialized structure, we convert the velocity FWHM to a velocity dispersion by the relation:
\begin{equation}
    \sigma_v = \frac{v_{\rm FWHM}}{2~\sqrt{{\rm ln ~2} }}.
\end{equation}
We then use this velocity dispersion to infer a virial radius, $R_{\rm 200}$, of the cluster:
\begin{equation}
    R_{200}^{\rm infer} =  \frac{\sigma_v}{4/3\pi G \Delta_{200}\rho_{\rm crit}},
\end{equation} 
where $\Delta_{200}$ and $\rho_{\rm crit}$ are the overdensity and critical density, respectively.  We then adopt new redshifts (solely for the purpose of feeding MCPM) about the median redshift of the cluster members by sampling from a normal distribution with standard deviation corresponding to the change in redshift that would result in a luminosity distance difference equal to the inferred $R_{\rm 200}$.  Finally, we convert these galaxy coordinates and adopted redshifts to 3D Cartesian space via luminosity distances based on the new redshifts; these then serve as inputs to MCPM.

\section{Data Products}

\subsection{The Catalog}
The final value-added catalog contains the positions and redshifts and the stellar mass of the galaxies in the NASA-Sloan Atlas and the eBOSS Firefly Value-Added Catalog. We include a column, \texttt{MASS\_SOURCE}, to indicate which catalog was used to estimate the mass. The MCPM algorithm uses the galaxy mass to build the matter density field. The primary field of interest here is \texttt{MATTERDENS}, the matter density field at the location of a given galaxy, which was derived from fits of MCPM models in 3D volumes and mapped to the cosmological matter density (relative to the mean matter density) using MCPM fits to the Bolshoi-Planck simulations. The \texttt{catalogID} is a combination of \texttt{plate-mjd-fiberid}. A unique identifier is the combination of \texttt{catalogID} and \texttt{mcpmRun}. Objects with the same value of \texttt{mcpmRun} were fitted with the MCPM model simultaneously. The data were sliced in redshift to yield samples producing self-consistent large-scale structures over the volume in each slice. \texttt{mcpmRun} = 0 correspond to $0.01 < z < 0.1$ SDSS galaxies with masses from the NASA/Sloan Atlas. Samples of LOWZ LRGs are marked 1-2 $(z < 0.20)$, 3-4 $(0.18 < z < 0.30)$, 5-6 $(0.28 < z < 0.38)$, and 7-8 $(0.36 < z < 0.51)$; each pair (e.g., 3-4), corresponds to the NGC/SGC samples in some redshift slice, with odd and even numbers for NGC and SGC, respectively. The data model for the catalog is described in Table~\ref{tab:columns}.

\begin{deluxetable*}{cccc}
\tablecaption{Data Model \label{tab:columns}}
\tablehead{\colhead{Name} & \colhead{Type} & \colhead{Unit} & \colhead{Description}
}
\colnumbers
\startdata
catalogID & char[13] &  & Combination of PLATE-MJD-FIBERID \\
plate & int32 &  & Plate number \\
mjd & int32 &  & MJD of observation \\
fiberid & int32 &  & Fiber identification number \\
ra & float64 & deg & Right ascension of fiber, J2000 \\
dec & float64 & deg & Declination of fiber, J2000 \\
z & float32 &  & Best redshift \\
massSource & char[7] &  & Source of the mass determination (nsa or firefly) \\
mcpmRun & int8 &  & Index of galaxy sample fitted simultaneously with MCPM \\
mstars & float64 & $M_{\odot}$ & Stellar mass \\
matterDens & float32 &  & log10 of the ratio of the matter density relative to the mean matter density \\
\enddata
\tablecomments{Schema for the MCPM Value-Added Catalog, v1.0.0 as found in \texttt{slimeMold\_galaxy\_catalog\_v1\_0\_0.fits}.}
\end{deluxetable*}

\subsection{3D Density Cube}

In addition to the VAC, which contains the density at the location of each galaxy, we offer the full 3D density field of the relevant volumes, available at \url{https://data.sdss.org/sas/dr17/eboss/lss/mcpm/v1\_0\_0/datacube/}. These may be queried using our custom package, \texttt{pyslime}\footnote{https://github.com/jnburchett/pyslime}. The data will unzip to a directory which may be opened by \texttt{pyslime}. This will enable the user to query the overdensity at arbitrary points in the cube, allowing the study of voids and filamentary structures outside the local environment of the input galaxy field.

\section{Discussion}
\subsection{Comparison to \cite{peng10}}

We can additionally validate our model by comparing our findings to that of other surveys. Although \cite{burchett20} has demonstrated the efficacy of our model, we present comparisons to other studies, leveraging our deeper and larger surveys. 

\begin{figure}
\includegraphics[scale=0.33]{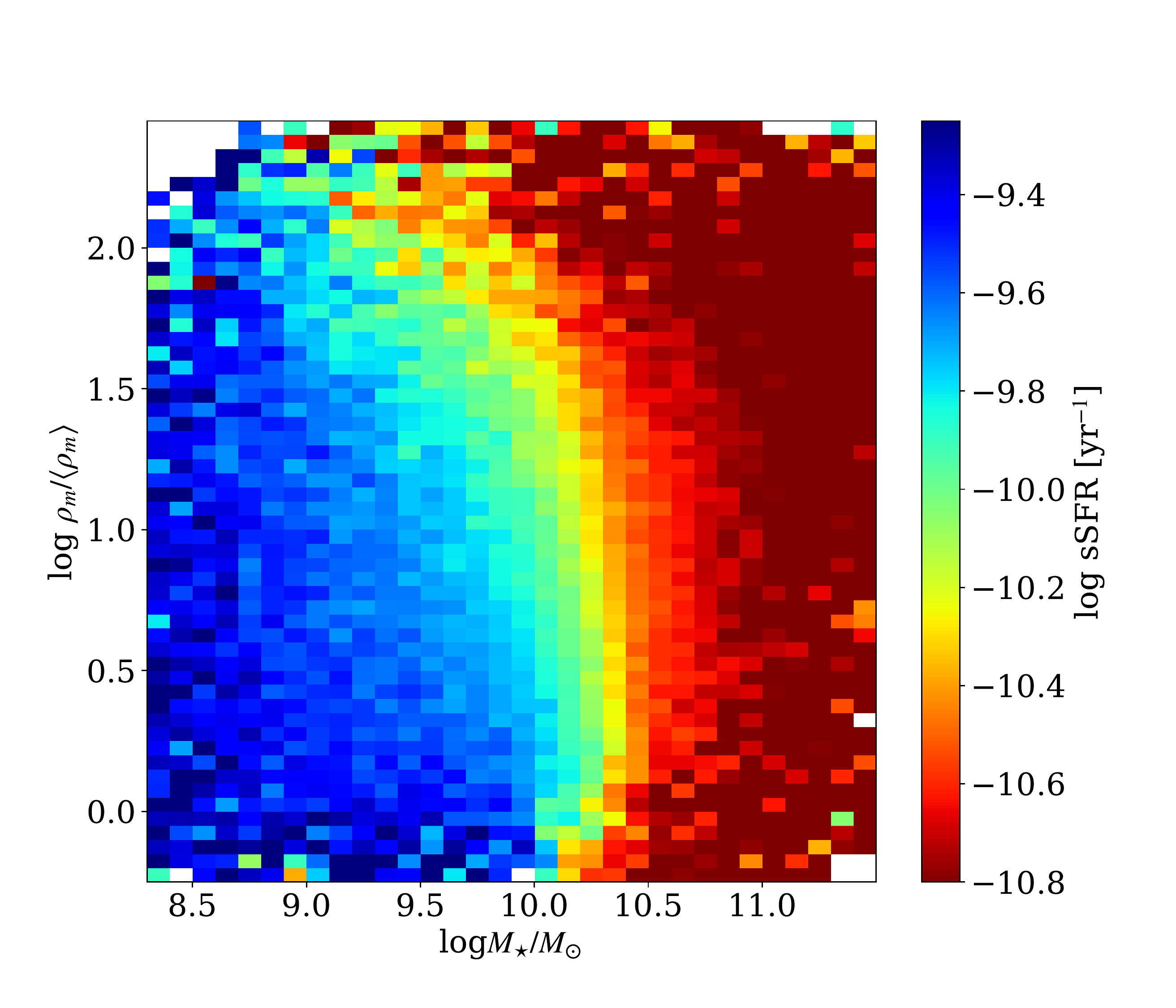}
\caption{The dependence of star formation activity on galaxy environment and stellar mass for the galaxies within the NSA/SDSS volume ($z < 0.1$). The color coding denotes sSFR in the population within each mass/environment bin, where the environmental density is determined from our MCPM cosmic web reconstruction algorithm. A comparison with Figure 6 of \cite{peng10} shows a similarly increasing red fraction as a function of both mass at fixed density and density at fixed mass.
~\\
\label{fig:peng}}
\end{figure}

\citet[]{peng10} used a method based on the 5th nearest galaxy neighbors to estimate the environmental density and studied the SFR and the quenched fraction of galaxies as a functions of this density metric and galaxy mass.  \cite{burchett20} illustrate that the MCPM method of computing cosmic density qualitatively matches the results \cite[see Figure 5 \& 6 in][]{peng10}. In Figure~\ref{fig:peng}, we demonstrate the improvement in signal gained with the NSA/SDSS sample as the increase in the number of galaxies is significant and the reproduction of their density-stellar mass-sSFR relations.
        
\subsection{Potential Applications}

Our primary aim in this manuscript is to showcase the dataset and describe its construction. There are, however, many exciting applications for this dataset that are well beyond the scope of this publication. Here, we list four general areas of application: 

\begin{itemize}
    \item \textbf{Galaxy evolution in the cosmic web:} A vast amount of galaxy properties measured and inferred from both multiwavelength photometry and spectroscopy have been cataloged for SDSS galaxies \citep[many also released as VACs; e.g.,][]{Salim:2016_GSWLC1} via straightforward crossmatching with our catalog, myriad galaxy-environment analyses may be readily conducted.  Figure~\ref{fig:peng} highlights one direct application of the galaxy-density catalog to study the possible impacts of a galaxy's location within the cosmic web on its evolution.  In particular, Figure~\ref{fig:peng} shows the dependence of star formation activity as a function of large-scale structure density.  Our dataset is ideal for comparing effects induced by the more local environment (groups/clusters) to those induced by the cosmic web. 
    
    \item \textbf{Void finding:} In the linear regime, the sizes of voids and their correlation statistics are sensitive to cosmology, particularly dark energy \citep{Pisani:2015_countingVoids}. Although most of the analyses we have alluded to thus far focus on the denser regions of the cosmic web, namely filaments and nodes, our density cubes naturally include the underdense regions.  Simple centroiding and clustering algorithms may be readily applied to these density fields to directly identify and characterize the voids, which in turn may be used as inputs for cosmological parameter estimations using, e.g., the Alcock-Paczynski effect \citep{AlcockPaczynski:1979_test}. 
    
    \item \textbf{The intergalactic medium:}  Hydrodynamical cosmological simulations predict a rich multiphase structure in the intergalactic gas permeating throughout the cosmic web \citep[e.g.,][]{Cen:1999yq,Dave:2001kx,Tepper-Garcia:2012aa}.  In addition to the physical states of gas resulting from large-scale structure formation \citep{Bertschinger:1985fk, Molnar:2009fk}, energetic feedback from the galaxies themselves might extend well beyond the virial radius, which is often adopted as a fiducial extent of a galaxy's halo \citep{Finlator:2008_MZR,Schaye:2015yg,Nelson:2019_outflows}. \citet{burchett20} used HST-observed background quasar sightlines through the MCPM reconstructed volume to find a relationship between cosmic web density and Ly$\alpha$ optical depth.  A similar analysis could and should be done leveraging our higher redshift LRG reconstruction with other absorption tracers, such as Mg II.   
    
    \item \textbf{Multimessenger transient followup:}  Transient phenomena, such as gravitational waves and fast radio bursts, are typically detected with imprecise localization, with scales of minutes or degrees on the sky \citep{Chen:2016_gwLocalization,CHIME:2019_eightRepeaters}. Space-based and ground-based facilities around the world then follow up these detections to identify and characterize the sources \citep[e.g.,][]{Coulter:2017_kilonova}. As extragalactic sources are statistically more likely to be found within the large-scale structure, transient observers could employ our reconstructed density field of the cosmic web in follow-up imaging campaigns to prioritize pointings toward regions of the sky most likely to contain the source counterparts.
\end{itemize}

\subsection{Known Limitations}

The VAC volumes have the usual luminosity function systematics that are present in the underlying SDSS and LRG catalogs. Specifically, the sampling density of galaxies is more significant at lower redshifts. This is reflected in the trace and can be seen in the SDSS data as well as each slice of the LRG catalogs, as shown in Figure~\ref{fig:slices-diagram}. This presents itself as an increased density at the lower redshift end of the volume. However, the mean matter density at the low and high redshift ends of each volume is consistent.

Some sub-optimality of the model fit arises from the fact that the optimal \param{sensing distance} grows linearly according to the data in Figure~\ref{fig:slices}, whereas the catalog is a piece-wise constant approximation of this.

Due to the differing \param{sensing distance} in each slice, there is a slight discontinuity of the MCPM densities extracted from the overlaps between the LRG slices. Thus, we recommend comparing densities on a slice-by-slice basis and avoiding comparing quantities based on the density at different redshift slices.

\section{Conclusion}
Herein we leverage the \textit{Monte Carlo Physarum Machine} (MCPM) methodology, inspired by the growth and movement of Physarum polycephalum slime mold, to map the cosmic web within several sub-samples of the SDSS spectroscopic galaxy catalogs. The MCPM model inputs a galaxy field with known masses and outputs the large-scale structure density field. We train our model using the Bolshoi-Planck cosmological survey, producing a reconstruction of the simulated cosmic web where the underlying density is known. Using the simulation as ground truth, we describe the supervised tuning of MCPM parameters to produce an optimal fit. We apply this tuned model to the NASA-Sloan Atlas and the eBOSS LRG Firefly Value-Added Catalogs to create both a 3D density cube and a catalog of cosmic densities at the location of the galaxies. The SDSS NASA-Sloan Atlas catalogs include a more complete galaxy sample at $z < 0.1$. We describe and employ a novel method on this dataset to reduce the effect of peculiar motions on the spectroscopic distances. The MCPM fits to the eBOSS LRG North and South Galactic Cap catalogs capture the larger-scale cosmic web out to $z \lesssim 0.5$. This paper describes the release the \textit{Cosmic Slime Value Added Catalog}, part of SDSS DR17, which is the combination the two galaxy catalogs with density estimates as well as the resultant 3D density cubes of the two galaxy samples.
Finally, we highlight some exciting potential applications of this data set, which include galaxy evolution in the context of the cosmic web, void finding, studies of the intergalactic medium, and multimessenger transient followup. 

\section{Acknowledgements}
The authors would like to especially acknowledge Joel Primack and Doug Hellinger for sharing the outputs of the Boloshoi-Planck simulations. We also gratefully acknowledge the hospitality and support of the 2019 Kavli Summer Program in Astrophysics at UC Santa Cruz.

JB would like to acknowledge funding support from the National Science Foundation LEAPS-MPS award \#2137452. 

MCW and JKW acknowledge support for this work from NSF-AST 1812521, NSF-CAREER 2044303, the Research Corporation for Science Advancement, grant ID number 26842.

OE is supported by an incubator fellowship of the Open Source Program Office at UC Santa Cruz funded by the Alfred P. Sloan Foundation (G-2021-16957).

Funding for the Sloan Digital Sky Survey IV has been provided by the Alfred P. Sloan Foundation, the U.S. Department of Energy Office of Science, and the Participating Institutions. SDSS-IV acknowledges
support and resources from the Center for High-Performance Computing at
the University of Utah. The SDSS website is www.sdss.org.

SDSS-IV is managed by the Astrophysical Research Consortium for the 
Participating Institutions of the SDSS Collaboration including the 
Brazilian Participation Group, the Carnegie Institution for Science, 
Carnegie Mellon University, the Chilean Participation Group, the French Participation Group, Harvard-Smithsonian Center for Astrophysics, 
Instituto de Astrof\'isica de Canarias, The Johns Hopkins University, Kavli Institute for the Physics and Mathematics of the Universe (IPMU) / 
University of Tokyo, the Korean Participation Group, Lawrence Berkeley National Laboratory, 
Leibniz Institut f\"ur Astrophysik Potsdam (AIP),  
Max-Planck-Institut f\"ur Astronomie (MPIA Heidelberg), 
Max-Planck-Institut f\"ur Astrophysik (MPA Garching), 
Max-Planck-Institut f\"ur Extraterrestrische Physik (MPE), 
National Astronomical Observatories of China, New Mexico State University, 
New York University, University of Notre Dame, 
Observat\'ario Nacional / MCTI, The Ohio State University, 
Pennsylvania State University, Shanghai Astronomical Observatory, 
United Kingdom Participation Group,
Universidad Nacional Aut\'onoma de M\'exico, University of Arizona, 
University of Colorado Boulder, University of Oxford, University of Portsmouth, 
University of Utah, University of Virginia, University of Washington, University of Wisconsin, 
Vanderbilt University, and Yale University.

\bibliography{main.bbl}
\end{document}